\documentclass[usenatbib,onecolumn,useAMS]{mn2e}
\usepackage{epsfig}

\def\msun{{\rm M_{\odot}}}

\title[Kinetic theory viscosity] {Kinetic theory viscosity} 
\author[C.J.~Clarke, J.E.~Pringle]{C.J.~Clarke$^1$, J.E.~Pringle$^{1,2}$\\
$^1$Institute of Astronomy, Madingley Rd, Cambridge, CB3 0HA, UK\\
$^2$Space Telescope Science Institute, 3700 San Martin
Drive, Baltimore, MD 21218, USA}
\date{Submitted: July 2103}

\pagerange{\pageref{firstpage}--\pageref{lastpage}} \pubyear{2003}

\begin{document}
\def\lta{\mathrel{\spose{\lower 3pt\hbox{$\mathchar"218$}}
     \raise 2.0pt\hbox{$\mathchar"13C$}}}
\def\gta{\mathrel{\spose{\lower 3pt\hbox{$\mathchar"218$}}
     \raise 2.0pt\hbox{$\mathchar"13E$}}}
\def\Msun{{\rm M}_\odot}
\def\msun{{\rm M}_\odot}
\def\Rsun{{\rm R}_\odot}
\def\Lsun{{\rm L}_\odot}
\def\19{GRS~1915+105}
\label{firstpage}
\maketitle

\begin{abstract}

We show how the viscous evolution of Keplerian accretion 
discs can be understood in terms of simple kinetic theory. Although 
standard physics texts give a simple derivation of momentum transfer
in a linear shear flow using kinetic theory, 
many authors, as detailed by Hayashi \& Matsuda 2001,
have had difficulties applying the
same considerations to a circular shear flow. We show here how this
may be done, and note that the essential ingredients are to take
proper account of, first, isotropy locally in the frame of the fluid
and, second, the geometry of the mean flow.

\end{abstract}


\section{Introduction}

Accretion discs play a central role in a wide range of astronomical
environments, mediating the gas flows in the vicinity of object as
diverse as AGN, binary stars and protostars (Pringle, 1981). In an
accretion disc, the predominant flow is a circular shear flow, with
angular velocity $\Omega(R)$ a function of radius $R$ from the central
object. Accretion takes place because of the action of some form of
dissipation which releases the free energy of the shear flow as heat,
and so allows the disc material to fall deeper into the potential well
of the central object. Simple physical energy arguments ({\it e.g.}
Lynden-Bell \& Pringle, 1974) indicate that the dissipative process
must take the form of a stress which transports angular momentum
outwards. Because the free energy of a circular shear flow is zero
only if $d\Omega/dR = 0$, it follows that the relevant element of the
stress tensor must be of the form
\begin{equation}
T_{\phi R} \propto - d\Omega/dR,
\end{equation}
where $\phi$ is the azimuthal coordinate.

This can be deduced from the standard derivation of Navier-Stokes
stress to be found in the fluid dynamics textbooks. However, in a
recent paper, Hayashi \& Matsuda (2001) have drawn attention to the
fact that attempts to provide a physical explanation of the above
result in terms of simple kinetic theory have resulted in failure. 

The simple kinetic explanation given in basic physics text books for
the effect of viscosity on a simple linear shear flow, of the form
${\bf u} = (0,U(x))$ in Cartesian coordinates, relies on the fact that
the kinetic particles conserve linear momentum between
collisions. Thus particles crossing some fiducial plane $x = x_o$ tend
to mix up and smooth out the momentum distribution of the fluid 
and so 
give rise to a stress of the form $T_{yx} \propto -dU/dx$.

However, as Hayashi \& Matsuda (2001) point out, it is in the
application of these simple concepts to a circular shear flow that the
problems seem to arise. The simple generalisation that, in a circular
shear flow, the kinetic particles conserve angular momentum $j = R^2
\Omega$ between collisions would imply, taken at face value, that the
the movement of particles across some fiducial circle $R=R_o$ would
tend to try to mix up and smooth out the distribution of angular
momentum of the fluid, and thus that the stress would be proportional
to $-dj/dR$ (see e.g. Madej \& Paczynski 1977). From the arguments given above, this is clearly
wrong. Not only would this predict a stress in the case when the
shear $d\Omega/dR$ is zero, but for a standard Keplerian accretion
disc for which $j \propto R^{1/2}$ it would transport angular momentum
inwards rather than outwards. As detailed by Hayashi \& Matsuda (2001)
attempts to get round this and to produce the 'correct' result have
only succeeded by making mathematical errors in the derivation.

From all these problems, Hayashi \& Matsuda (2001) conclude that
although what they call the derivation 'with mathematical rigour'
({\it i.e.} the usual Navier-Stokes argument) gives the correct
answer, in order to obtain the correct answer using kinetic theory one
must take account of such complications as Coriolis force.  In this
paper, we shall show that, although it is obviously necessary to
include Coriolis force if one works in a frame co-rotating with the
flow, one can obtain the correct result from straightforward kinetic
theory in the inertial frame.

Before we do so, it is instructive to return to the 'mathematical'
relationship between stress and strain derived in the standard fluid
textbooks (see, for example, Batchelor, 1967, Section 3.3). The basic
point we wish to make is that the standard Navier-Stokes expression
for momentum transfer ({\it i.e.} stress) is based on a simple {\it
physical} argument. The argument may involve the use of tensors, which
physicists tend to meet in courses on mathematical methods, but the
argument itself is not a mathematical one. The point is that stress
(momentum transfer) in a fluid (or in a solid) can be expressed as a
second order tensor. This is a physical (and not a mathematical)
statement, in exactly the same sense that the statement that a
velocity is a first order tensor (i.e. a vector) is a physical, and
not a mathematical, statement.\footnote{The mathematical complications
come in when one has to specify a particular tensor (or vector) by
specifying the coordinate representation of that tensor in a
particular coordinate frame, and get worse when one has to specify the
representation of that same tensor in some other coordinate frame.}
For a fluid, the physical {\it Ansatz} is simply that the stress
tensor must be physically related to the rate of strain tensor (which
is a second order tensor which derives from the first derivatives of
the velocity field, and thus incorporates information about the
shear). The simplest assumption is that relationship between these two
tensors is through a (fourth order) tensor which is isotropic. It is
this assumption of the isotropy of the relationship, which is based on
the physical assumption that the fluid itself is isotropic, which gives
rise to the standard Navier-Stokes expression for the
viscosity.\footnote{This simplest assumption is not necessarily the
correct one. For example it is possible that the viscosity might
depend on the absolute orientation of the shear in some inertial frame
({\it e.g.}  the stress due to convection in a rotating medium -- see
Kumar, Narayan \& Loeb 1995, and references therein; the stress due to
non-isotropic mixing -- Bretherton \& Turner, 1968; or the stress
induced by the magneto-rotational instability -- Torkelsson {\it
et. al.} 2000).}

It is important to realise that these tensors ({\it i.e.} scalars,
vectors, second order tensors) exist as physical quantities,
independent of any coordinate system. The statement that there is a
relationship between two of them is a physical statement. It is only
when one has to calculate a particular element of the stress tensor,
for example corresponding to linear momentum transfer in a linear
shear flow, or to the angular momentum transfer in a circular shear
flow, that one has to evaluate coordinate dependent expressions, which
can get mathematically complicated. But when one does this, one finds
that the flux of linear momentum in a linear shear flow just depends
on $-dU/dx$, and that the angular momentum flux in a circular shear
flow just depends on $-d\Omega/dR$. However, this result also enables
us to draw another conclusion. The reason for the difference between
the terms $dU/dx$ and $d\Omega/dR$ is due solely to the difference
between the coordinate systems. That is, it comes from geometry
alone. This implies that when looking for differences in derivations
for simple kinetic theory formulae between the linear and circular
shear flows, we need only concern ourselves with geometry, and not
with dynamical complications such as Coriolis force. In addition, we
also need to take note of the fact that the Navier-Stokes expression
does depend critically on assumptions about isotropy of the
fluid. Thus we should expect to have to make a similar assumption
about the properties of our kinetic particles.

\section{Kinetic Theory}

In this Section we compute the relationship between stress and (rate
of) strain for two simple shear flows using kinetic theory. To keep
the concepts and the algebra simple, we make a number of simplifying
assumptions. We work in two dimensions only. That is, we assume that
the kinetic particles (assumed to be identical, with mass $m$) move
only in a two-dimensional plane.  We assume that the net effect of the
scattering processes within the fluid is that these particles are
emitted at a constant rate at each point of the fluid, and that each
particle is emitted with an identical velocity, $c$, relative to the
local fluid.\footnote{It is of course straightforward to generalise
the analysis to take account of the particles being emitted with a
distribution of velocities.}.  We represent by $\dot N(\lambda) $ the
number of particles emitted per unit area per unit time that travel a
distance greater than $\lambda$ before colliding with another particle
and we assume that $\dot N(\lambda) $ is independent of position.  The
requirement of isotropy, as discussed above, can now be imposed by
making the assumption that this emission takes place at each place in
the fluid isotropically {\it in the frame of the fluid} at that place.

\subsection{Plane shear flow}

In Cartesian $(x,y)$ coordinates we let the background fluid flow be
of the form

\begin{equation}
{\bf u} = ( 0 , U(x)).
\end{equation}

We consider the stress acting on a line element, length $dl$, centred
at point S, at position $ ( x_0, 0 )$, and lying in the $y-$direction,
that is, with unit normal in the $x-$direction.  For a linear shear
flow, we consider the case

\begin{equation}
U(x) = U_0 +(x-x_0)U^\prime, 
\end{equation}
where $U^\prime \equiv dU/dx$ is a constant. Note that although we
consider here a linear shear flow with constant shear, any linear
shear flow can be treated as a constant shear for $x \approx x_0$ and
with $U^\prime = dU/dx$, evaluated at $x = x_0$.

\begin{figure}
\vspace{2pt}
\epsfig{file=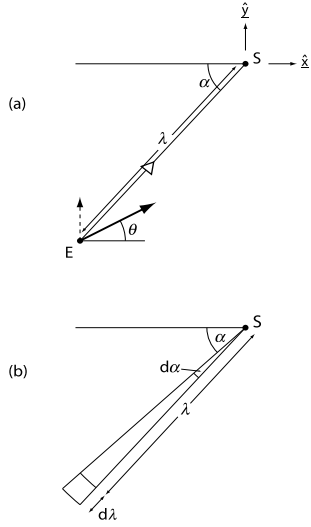,width=8.5cm,height=7.cm}
\caption{(a) Schematic of linear shear flow showing emission point, E,
line element S and path of particles, ES, in frame of S. The dashed
arrow denotes the velocity of of the mean flow at E with respect
to that at S, the bold arrow the random velocity component and the
open arrowhead denotes the particle resultant velocity in the frame
of S. Note that in reality (although not in the Figure) the length
of the bold arrow is much greater than that of the dashed arrow.
(b) Schematic of linear shear flow showing the patch around E that
is the area element for integration of the viscous stress.
}
\end{figure}

For convenience, we work in the frame comoving with the mean fluid
flow at S. This frame is an inertial frame, and in this frame the
mean flow is given by

\begin{equation}
{\bf u} = ( 0 ,  (x-x_0) U^\prime).
\end{equation}
 
We now focus on particles that are emitted from a point E that is
located at a distance $\lambda$ from S and where the line SE makes an
angle $\alpha$ with the negative $x-$axis (see Figure 1a). The
critical point is that {\it in the frame of the fluid at }E the
particles that travel along ES are emitted at an angle $\theta$ to the
$x-$axis where the relationship between $\theta$ and $\alpha$ is
simply deduced by considering the ratio of the $x-$ and $y-$components
of the particle velocity in the rest frame of S, {\it i.e.}

\begin{equation}
\tan \alpha = {{c \sin \theta - U^\prime \lambda \cos \alpha}\over{c \cos \theta}}.
\end{equation}

We consider the limit that the shear velocity across a typical mean
free path is much less than the random particle velocity ({\it i.e.}
$U^\prime \lambda \ll c$), so that (to O$(U^\prime \lambda /c)$) we
may write

\begin{equation}
\label{tantheta}
\tan \theta \sim \tan \alpha + U^\prime \lambda/c,
\end {equation}

\begin{equation}
\cos \theta \sim \cos \alpha - \sin \alpha \cos^2 \alpha \, U^\prime \lambda/c, 
\end {equation}
and
\begin{equation}
\label{sintheta}
\sin \theta \sim \sin \alpha + \cos^3 \alpha \,U^\prime \lambda/c. 
\end {equation}

In order to compute the rate of arrival of $y$-momentum at S due to
particles originating near E, we consider a patch around E that is a
portion of an annulus centred on S (radius $\lambda$, thickness
$d\lambda$) where the patch subtends an angle $d \alpha$ at S (see
Figure 1b). The rate of emission of particles from this patch that
travel far enough before colliding to be able to reach S is simply
$\dot N(\lambda) \lambda d\lambda d \alpha$. Such particles emanate
isotropically from E {\it in the rest frame of the fluid at }E. Thus
the fraction, $f$, of such particles that impinge on the line element
$dl$ at S is

\begin{equation}
f = {{dl \cos \alpha}\over{2 \pi L}},
\end{equation}
where $L$ is the distance traveled by the particles between E and S
{\it in the rest frame of the fluid at }E. The relationship between
$L$ and $\lambda$ is simply ascertained by noting that the relative
velocity between E and S is zero in the $x-$direction and hence that
the distances travelled in the $x-$direction between E and S in the
two frames is thus equal (i.e. $L \cos \theta = \lambda \cos \alpha$)
from which we deduce, to first order in $\lambda$,

\begin{equation}
L \sim \lambda (1 - \sin \alpha \cos \alpha U^\prime \lambda/c)^{-1},
\end{equation}
and thus  that

\begin{equation}
\label{f}
f \sim  {{dl \cos \alpha(1 - \sin \alpha \cos \alpha U^\prime \lambda/c)}\over{2 \pi \lambda}}.
\end{equation}
 
Finally, the $y-$velocity of each particle emitted from around E is

\begin{equation}
v_y = -U^\prime \lambda \cos \alpha + c \sin \theta, 
\end{equation}
which (using equation~\ref{sintheta}) may be approximated as

\begin{equation}
\label{vy}
v_y \sim c \sin \alpha - U^\prime \lambda \cos \alpha \sin^2 \alpha.
\end{equation}

Thus the total rate of arrival of $y$-momentum at S from the patch at
E is 

\begin{equation}
\dot p_y = m v_y f \dot N(\lambda) \lambda d\lambda d \alpha.
\end{equation}

Substituting for $f$ and $v_y$ from equations~\ref{f} and~\ref{vy} we may
obtain, to the order of our approximation, the total rate of arrival
of $y-$momentum at S by integrating over $\alpha$ and $\lambda$, in the
form

\begin{equation}
\dot p_y \sim \int_{0}^{\infty} {{m \dot N(\lambda) dl }\over{2 \pi}} \int_{-\pi/2}^{\pi/2} \cos \alpha (1 - \sin \alpha \cos \alpha U^\prime \lambda/c)(c \sin \alpha - U^\prime \lambda \cos \alpha \sin^2 \alpha) d\alpha  \, d\lambda. 
\end{equation}
Since we are only
considering the momentum flux due to particles arriving from one side
of the line element (that is, particles with $x < x_o$), then 
 $\alpha $ runs from $-\pi/2$ to $\pi/2$.

\begin{equation}
\label{pydot}
\dot p_y \sim -{{m U^\prime dl}\over{8}} \int_{0}^{\infty} \dot N(\lambda) \lambda d\lambda.
\end{equation}

[We note that the simple arguments found in many physics
text books for  kinetic theory viscosity in a linear shear flow, do 
{\it not} correctly take account of the fact that the particle
velocity distribution is isotropic in the frame of the fluid at the point
of {\it emission} (i.e. last collision), 
rather than at the point where the momentum flux
is measured (see for example, Jeans 1940). In a linear
shear flow, this  error only changes the
coefficient in front of the viscosity coefficient and retains the 
correct form
of the equation for the resulting viscous stress. In the circular
case, however, we shall find that it is necessary to treat this
subtlety correctly, even  in order to obtain the  correct
functional form of the viscous stress].

\subsection{Circular shear flow}

We now carry out essentially the same analysis, but this time for a
circular shear flow. We still work in an inertial frame, (that is, a
frame in which the particle trajectories are straight lines) but in
this case the underlying fluid flow is, in cylindrical polar
coordinates $(R,\phi)$, of the form
\begin{equation}
{\bf u} = (0, R\Omega(R)).
\end{equation}

We consider the stress acting on a (small) line element, length $dl$,
centred at point S which in Cartesian coordinates, centred at the
origin O ({\it i.e.} at $R=0$), lies at $(R_0, 0)$. The line element
lies in the azimuthal direction, and so has unit normal in the radial
$R-$direction.  As before, we work in the {\it inertial} frame that
co-moves with the mean fluid flow at S and again consider the limit
that the shear across a mean free path $\lambda$ is much less than the
random velocity $c$ ({\it i.e.} $ R \Omega^\prime \lambda \ll
c$).\footnote{For a Keplerian accretion disc the disc thickness is $H
\sim c/\Omega$, and thus this approximation implies that $\lambda \ll
H$. Thus our analysis does not cover astrophysical situations in which
the mean free path is comparable to, or larger than the disc thickness
(Paczynski, 1978; Brahic, 1977)} We note that, since $ R \Omega^\prime
\sim \Omega$, it is also the case that $\lambda \ll R$. We defer
until the end of this Section 
a discussion of the constraints placed on $\lambda$ by our neglect of the
curvature of particle orbits between collisions.   

\begin{figure}
\vspace{2pt}
\epsfig{file=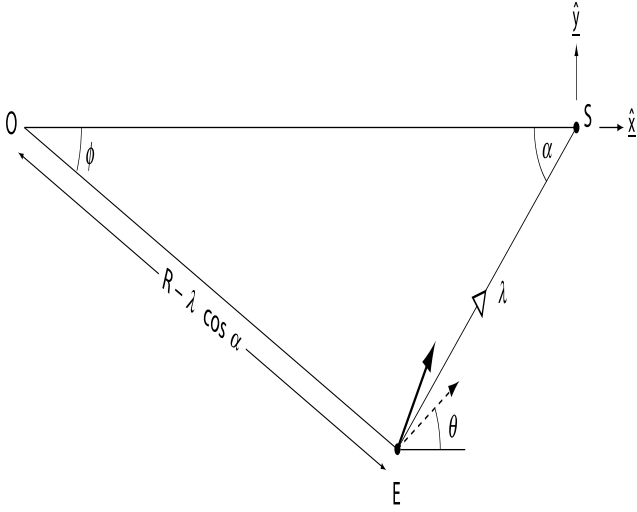,width=8.5cm,height=7.cm}
\caption{ Schematic of circular shear flow. Symbols as in Figure 1.
Note that the dashed arrow (the velocity of the mean flow at
E relative to that at S) now has a component in the x direction due
to the geometry of the mean flow}.
\end{figure}

As before, we consider particles that are emitted in the vicinity of a
point E, located at distance $\lambda$ from S where ES makes an angle
$\alpha $ with the inward directed radius vector at S.  The only
difference from the foregoing analysis is that since the streamlines
are now circular, the velocity of the mean flow at E with respect to
that at S has non-zero components in both the $x-$ and $y-$directions
(see Figure~2). Specifically, the point E is located at radius $R_{\rm
E} = R_0- \lambda \cos \alpha$ (to O$(\lambda/R)$) where the velocity
of the mean flow (with respect to the origin) is in the azimuthal
direction and of magnitude $ R_0 \Omega_0 - \lambda \cos \alpha
(\Omega_0 + R_0 \Omega^\prime)$ ($\Omega_0$ and $\Omega^\prime$ being
respectively the angular velocity and its gradient evaluated at
S). The radial vector at E makes an angle

\begin{equation}
\phi \sim {{\lambda}\over{R}} {{\sin \alpha}}
\end{equation}
with the $x-$axis. Consequently the $x-$ and $y-$components of the mean flow
velocity at E with respect to a frame co-moving with the mean
flow at S are (to O$(\lambda/R)$) 

\begin{equation}  
\label{vxrel}
v_{x}^{\rm rel} \sim {{\lambda \Omega_0 \sin \alpha}},
\end{equation}
and

\begin{equation}
v_{y}^{\rm rel} \sim - \lambda \cos \alpha (\Omega_0 +
R_0 \Omega^\prime).
\end{equation}

As before, we can establish the relationship between the angles $\alpha$
and $\theta$ using

\begin{equation}
\tan \alpha = {{v_{y}^{\rm rel}+ c \sin \theta}\over{v_{x}^{\rm rel} + c \cos \theta}},
\end{equation}
so that, in this case,

\begin{equation}
\tan \theta \sim \tan \alpha + (\Omega_0 +
R_0 \Omega^\prime) \lambda /c  +
\lambda \Omega_0 \tan^2 \alpha /c,
\end{equation}

\begin{equation}
\cos \theta \sim \cos \alpha  - \sin \alpha \cos^2 \alpha
(\Omega_0 +
R_0 \Omega^\prime)\lambda /c  -  \lambda \Omega_0   \sin^3 \alpha /c,
\end{equation}
and
\begin{equation}
\sin \theta \sim \sin \alpha + \cos^3 \alpha (\Omega_0 +
R_0 \Omega^\prime) \lambda /c + \lambda \Omega_0 \sin^2 \alpha \cos \alpha /c.
\end{equation}

We note the similarity of these three equations to
equations~\ref{tantheta} --~\ref{sintheta}, since in the circular flow
$U^\prime$ corresponds simply to $(\Omega_0 + R_0 \Omega^\prime)$. In
each equation there is, however, an additional term (proportional to
$\Omega_0$) which takes care of the fact that the mean flow is not
plane-parallel in this case, and that hence each emission point has a
non-zero velocity with respect to S along the radial vector at S (see
equation~\ref{vxrel}). This modification in the relationship between
$\alpha$ and $\theta$ is crucial in explaining the different form of
the viscous stress in the circular case.

We also need to modify the relationship between $L$ and $\lambda$
compared with what we had previously, which we can do most simply by
equating the time of flight ES ($L/c$) with the time ($\lambda \cos
\alpha/(c \cos \theta + v_{x}^{\rm rel})$) required to traverse the
distance $\lambda \cos \alpha$ that separates E and S in the
$x-$direction.  After some algebra we obtain the expression

\begin{equation}
L \sim \lambda (1 - \sin \alpha \cos \alpha 
R_0 \Omega^\prime \lambda/c )^{-1},
\end{equation}
and hence find in this case that

\begin{equation}
\label{fcirc}
f \sim  {{dl \cos \alpha(1 - \sin \alpha \cos \alpha 
R_0 \Omega^\prime \lambda/c)}\over{2
 \pi \lambda}}.
\end{equation}

Finally, the $y-$velocity of each particle emitted from
E is 

\begin{equation}
v_y =c \sin \theta + v_{y}^{\rm rel},
\end{equation}
and hence, to the order to which we are working,

\begin{equation}
\label{vycirc}
v_y \sim c \sin \alpha - \sin^2 \alpha \cos \alpha \, R_0 \Omega^\prime \lambda
\end{equation}

Comparison of equations~\ref{f} and~\ref{vy} with equations~\ref{fcirc}
and ~\ref{vycirc} shows that the expressions are identical except in
as much as $U^\prime$ in the linear shear flow is replaced by $R_0
\Omega^\prime$ in the circular case. Thus (following the same
procedure for finding $\dot p_y$ as previously -- that is, by integrating
contributions to the $y$-momentum flux from all points with $R< R_0$) we
obtain (by analogy with equation~\ref{pydot}):
 
%\begin{equation}
%\dot p_y \sim -{{m R \Omega^\prime dl}\over{8}} \int_{0}^{\infty} \dot N(\lambda) \lamb
%da d\lambda
%\end{equation}

\begin{equation}
\label{pydotcirc}
\dot p_y \sim -{{m R \Omega^\prime dl}\over{8}} \int_{0}^{\infty} \dot N(\lambda) 
\lambda d\lambda
\end{equation}

We note that this analysis recovers the `correct' answer (as given by
the usual Navier-Stokes argument) that whereas the viscous stress in a
linear shear flow depends on the velocity gradient, in the case of a
circular flow the viscous stress instead depends on the gradient of
{\it angular} velocity.

  Finally, we turn to the issue of our neglect of the  curvature
of particle orbits between collisions. This neglect is always justified,
for example, in the case of laboratory Couette flow, where the fluid is not
subject to external long range forces. In the case that the 
acceleration of the mean flow {\it is}  provided by a long range central
force ({\it i.e.} by a force that is experienced by the particles between
collisions, as in the case of a Keplerian accretion disc) then the
change in ($x$-) velocity of a particle between collisions is $\delta v \sim
R \Omega^2 \lambda/c$. If $\delta v \ll c$ we can repeat the above analysis
by adding $\delta v$ to $v_x^{rel}$ (equation 19) and hence modifying the
subsequent equations relating $\theta$ to $\alpha$. We find that 
(to O($\delta v/c)$) the relation between $L$ and $\lambda$ (equation
(25)) and hence the expression for $f$ (equation (26)) is unchanged
by this addition but that $v_y$ (equation (28)) now contains
an additional term ($\delta v \sin^3 \alpha/c \cos \alpha$). This
term however makes a zero contribution to the $y$-momentum flux, when
integrated over $\alpha $, since the contributions from $\pm \alpha$
cancel. We thus find that our analysis is independent of the nature
of the central force providing the acceleration of the mean flow ,
provided that $\delta v \ll c$. \footnote{Note that it is {\it not}
necessary that $\delta v$ be much less than the shear velocity across
a particle mean free path $\lambda \Omega_0$,  
and that indeed this latter condition
{\it cannot} be met in a thin Keplerian accretion disc. We however find that
the symmetry of the terms in the momentum flux involving
$\delta v$  ensures that their contribution can be neglected, compared
with those involving $\lambda \Omega_0$, even when
$\delta v \gg \lambda \Omega_0$}. This requirement translates into
a condition on $\lambda$ for a thin Keplerian disc of the form
$\lambda \ll H^2/R$, which is more stringent than the condition
($\lambda \ll H$) derived above in order to justify our approach
of expanding quantities to first order in $\lambda \Omega_0/c$.

\section{Discussion}

We have shown that it is possible to obtain the correct expression for
the momentum transfer in a both a linear and a circular shear flow
using simple kinetic theory.  This correct expression implies that an
isotropic viscosity transfers momentum {\it down} an angular velocity
gradient. We have noted that it is essential to take proper account of
both the geometry and the fluid's isotropy.

 \begin{figure}
\vspace{2pt}
\epsfig{file=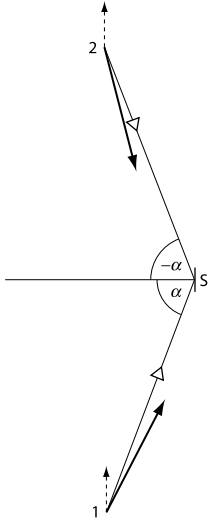,width=8.5cm,height=7.cm}
\caption{Velocity components for two particles, $1$ and $2$, in
a linear shear flow originating at emission points located at
$ \pm \alpha$. Key as in Figure 1.}
\end{figure}

Our analysis allows us to understand the flaw in the simple heuristic
argument applied to a circular shear flow (described in Section 1), which
leads to the conclusion that since angular momentum is conserved along
particle trajectories, it is the gradient in angular momentum which
particle mixing tends to smooth out. Particles arrive at the reference
patch, at point S, with momentum along the streamline ($y-$momentum)
that derives from two sources -- (i) the $y-$component of the random
emission velocity (assumed isotropic in the local rest frame of the
emission point) and (ii) the $y-$component of the streaming velocity
of the emission point relative to the reference patch.

 \begin{figure}
\vspace{2pt}
\epsfig{file=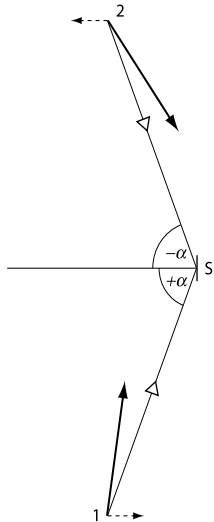,width=8.5cm,height=7.cm}
\caption{Velocity components for two particles, $1$ and $2$, in
a circular shear flow originating at emission points located at
$ \pm \alpha$. Key as in Figure 1.In this example, $R \Omega^\prime +
\Omega =0 $ (i.e. the speed of the flow is everywhere constant) but
nevertheless the total velocity vector (open arrowhead)
is larger for particle $1$, implying that there is a net flux of
$y$ momentum
at S.}
\end{figure}

The usual argument given to explain the momentum transfer in a {\it
linear} shear flow is simply that (ii) has the same sign for all the
emission points on one side of the reference patch and so that
momentum is added to, or subtracted from, the reference patch just
depending on the sign of the  velocity gradient in the mean
flow. In fact, there is another contribution, equal in magnitude and
sign to that described above, which may be understood by considering
the particles that are incident at the reference patch from $\pm
\alpha$. From Figure 3 it may be seen that the $x-$velocity (and hence
flux) of particles on the $+ \alpha$ side exceeds that on the $-
\alpha$ side ({\it i.e.} there is a greater flux of particles at the
reference patch whose emission is prograde than retrograde). Since the
$y-$momentum from (i) far exceeds that from (ii), this slight
asymmetry in the fluxes of prograde and retrograde particles produces
a net momentum flux that is equal to the more obvious source of
momentum flux described above.

We may now apply the same considerations to the {\it circular} shear
flow, where once again there are the two contributions to the
$y-$momentum -- (i) and (ii) described above. The important difference
now is that the mean streaming velocity is no longer in the
$y-$direction. The geometry of the circular arc ensures that this
contributes a positive (negative) $x-$velocity for $\alpha > 0 \:
(\alpha < 0)$, respectively. As may be seen from Figure 4, particles
that are incident from $+ \alpha$ (prograde particles) have a larger
amplitude of both $v_x$ and $v_y$ than those from $- \alpha$. This
ensures that the mean $y-$velocity of particles arriving at the
reference patch from a particular emission streamline is not equal to
the mean $y-$velocity of particles on that emission streamline. This
disparity is because of the relative boost in the arrival flux of
particles whose random velocity is prograde in the frame of the
emission streamline, compared with those that are retrograde.

This now allows us to understand the behaviour of a Keplerian
accretion disc at a qualitative level. The mean angular momentum of
particles orbiting at radii less than that of the reference patch is
less than that of the reference patch. If the distribution of angular
momenta of the particles arriving at the reference patch merely
reflected the distribution of angular momenta of particles on their
parent emission streamlines, this would imply that the arrival of
particles at the reference patch from smaller radii should exert a
spindown torque. Instead, the relative boost in the {\it arrival rate}
of particles that are emitted in the prograde direction, ensures that
the average angular momentum of the particles arriving at the
reference patch exceeds the average at the parent streamline. In the
Keplerian case, this relative boost in the arrival rate of the
prograde particles is enough to reverse the sign of angular momentum
transfer. Thus, particles arriving at the reference patch from smaller
radii exert a spin up torque, as required.

\section*{Acknowledgments}
CJC gratefully acknowledges support from the Leverhulme Trust.  JEP
gratefully acknowledges continuing support from the STScI Visitors'
Program. We acknowledge useful discussions on this problem with Peter
Scheuer and David Syer. We are indebted to the referee, Takuya Matsuda,
for interesting comments and for pointing out several typographical
errors in the manuscript.

%\section{References}

%\begin{references}

\end{document}